\documentclass[12pt]{article}
\usepackage{graphicx}
\usepackage[bookmarksnumbered,colorlinks,plainpages]{hyperref}

\begin{document}

\title{A mathematical method for irregular hamiltonian systems}

\author{Q.A. Wang and Wei Li\\
{\it Institut Sup\'erieur des Mat\'eriaux et M\'ecaniques Avanc\'es}, \\
{\it 44, Avenue F.A. Bartholdi, 72000 Le Mans, France}}

\date{}

\maketitle

\begin{abstract}
We present certain mathematical aspects of an information method which was
formulated in an attempt to investigate diffusion phenomena. We imagine a regular
dynamical hamiltonian systems under the random perturbation of thermal (molecular)
noise and chaotic motion. The random effect is taken into account via the
uncertainty of irregular dynamic process produced in this way. This uncertainty due
to different paths between two phase points is measured by a path information which
is maximized in connection with the action defined originally for the unperturbed
regular hamiltonian systems. The obtained transition probability depends
exponentially on this action. The usefulness of this information method has been
demonstrated by the derivation of diffusion laws without the usual assumptions. In
this work, some essential mathematical aspects of this irregular dynamics is
reviewed. It is emphasized that the classical action principle for single least
action path is no more valid and the formalism of classical mechanics for regular
hamiltonian systems is no more exact for irregular hamiltonian dynamics. There is
violation of the fundamental laws of mechanics by randomly perturbed hamiltonian
systems. However, the action principle is always present for the ensemble of paths
through the average action. This average action principle leads to a formalism of
stochastic mechanics in which, in spite of the violation of fundamental laws, the
mathematical form of classical mechanics can be recovered by a consideration of the
statistical averaging of the dynamics.
\end{abstract}

(classical mechanics)

\section{Introduction}
The dynamics of a Hamiltonian system can be roughly classified into two categories:
regular and irregular. The mathematics of the regular dynamics can be perfectly
formulated on the basis of the least action principle in classical mechanics. But the
stochastic character of irregular dynamics makes it much more complicated to be
formulated. Description of random dynamic process with Newtonian type equations is
possible if one introduces random forces like in Langevin equation\cite{Kubo}. Other
statistical method by informational consideration without introducing the details of
the dynamic randomness are proved to be very useful. The description of the stochastic
behavior of chaotic systems using Kolmogorov-Sinai entropy is an
example\cite{Dorfman}.

The method we address in this paper is originally intended to study diffusion process
by a informational consideration. Diffusion is an irregular dynamic process in which
components of a mixture are transported around the mixture by means of random
molecular motion. Over 200 years ago, Berthalot postulated\cite{Berthalot} that the
flow of mass by diffusion across a plane, was proportional to the concentration
gradient of the diffusant across that plane. About 50 years later, Fick
introduced\cite{Fick} two differential equations that quantified the above statement
for the case of transport through thin membranes. Fick's First Law states that the
flux $J$ of a component of concentration $n$ across a membrane is proportional to the
concentration gradient in the membrane:
\begin{eqnarray}                                            \label{w1}
J(x)=-D\frac{\partial n(x)}{\partial x}
\end{eqnarray}
where $x$ is the position variable for one dimensional systems. Fick's Second Law
states that the rate of time change of concentration of diffusant at a point is
proportional to the rate of spacial change of concentration gradient at that point
within the mixture
\begin{eqnarray}                                            \label{w2}
\frac{\partial n}{\partial t} = \frac{\partial}{\partial x}\left[D\frac{\partial
n(x)}{\partial x}\right].
\end{eqnarray}
If $D$ is constant everywhere in the mixture, the above equation becomes
$\frac{\partial n}{\partial t} = D\frac{\partial^2 n(x)}{\partial x^2}$.

Above normal diffusion laws are very precisely tested in experiments in most of
solids, liquids and gases and are widely studied in nonequilibrium thermostatistics
together with the Fokker-Planck equation of diffusion probability, the Fourier law of
heat conduction and the Ohm's law of electrical charge conduction. Many efforts to
derive theoretically the above diffusion laws were concentrated on special models of
solids in which particles are transported. Other phenomenological derivations are also
possible if one supposes Brownian motion and Markovian process\cite{Kubo}, or
Kolmogorov conditions\cite{Zaslavsky}. Recently, an attempt is made to derive these
laws with a different method based on information theory. The method consists in
considering hamiltonian systems under the perturbation of thermal noise and chaotic
instability. The dynamic uncertainty of this perturbed dynamic process is measured by
a path information connected with different paths. In order to derive the transition
probability distribution of the irregular process, the path information is maximized
in connection with, {\it faute de mieux}, the action defined with the unperturbed
Hamiltonian. The diffusion laws is simply the differential equation of the transition
probability distribution in exponential of action. Although we can hardly talk about
the exact action of a system subject to random forces and that the action here is only
the $unperturbed$ action of the system, the least action principle is still present in
this approach through the average action. This description of irregular dynamics
underlies a probabilistic formalism of mechanics which is logically different from the
conventional one based on the action principle for single path of the unperturbed
hamiltonian systems. In what follows, we discuss some mathematical aspects of this
stochastic mechanics.

\section{Regular dynamics of hamiltonian systems}
A hamiltonian system is a mechanical system which satisfies the following
Hamiltonian equations\cite{Dorfman} :

\begin{eqnarray}                                            \label{1}
\dot{x_i}=\frac{\partial H}{\partial p_i} \;\;and \;\; \dot{p_i}=-\frac{\partial
H}{\partial x_i}, \;\;with \;\; i=1,2,...n
\end{eqnarray}
where $n$ is the number of degrees of freedom, $x_i$ is the coordinates,
$P_i=m\dot{x_i}$ the momenta and $H$ is the hamiltonian of the system. The solutions
of these equations are geodesics in phase space $\Gamma$ of $2n$ dimension, i.e., the
paths between a initial point $a$ and a final point $b$ whose action $A_{k}=\int_a^b
L_{k}(t)dt$ has a stationary (least action principle). From this action principle, we
can derive Euler-Lagrange equations\cite{Arnold}
\begin{eqnarray}                                            \label{c7x}
\frac{\partial}{\partial t}\frac{\partial L_{k}(t)}{\partial \dot{x}}-\frac{\partial
L_{k}(t)}{\partial x}=0
\end{eqnarray}
where $L_{k}=T-V$ (we suppose $n=1$ from now on) is the Lagrangian of the system along
a path $k$. If we consider the Legendre transformation $H=P\dot{x}-L_{k}$,
Eqs.(\ref{1}) can be easily derived from Eq.(\ref{c7x})\cite{Arnold}. This formalism
of mechanics underlies a completely deterministic character of dynamic process: if the
initial conditions are given, there is no uncertainty, no information, no probability
distribution associated with the future evolution of the systems. The geodesics are
determined for any time by the above equations plus initial conditions.

\section{A path information}
Now we imagine a hamiltonian system under the perturbation of thermal noise and
chaotic instability. This perturbation can be either internal or external to the
system. A hamiltonian system containing a large number of particles in random motion
is an example of internal perturbation. The dynamic instability of chaotic systems is
another example of internal perturbation leading to stochastic deviation from regular
behavior. The motion of a Brownian particle is an example of external perturbation due
to the random molecular motion around the particle. Under these perturbations, the
geodesics of regular dynamics will be deformed and fluctuate in such a stochastic way
that following exactly the evolution of each mechanical quantity (e.g., action) is
inconceivable. We call this dynamic process irregular since there is uncertainty about
the future evolution even if the initial condition is given. We will consider two
dynamic uncertainties:

\begin{enumerate}

\item Between any two phase space points $a$ and $b$, there may be different possible
paths (labelled by $k$=1,2,...w) each having a probability $p_k(b|a)$ to be followed
by the system. This is the uncertainty considered by Feynman in his formulation of
quantum mechanics\cite{Feynman}. Here we introduce it within randomly perturbed
classical mechanics.

\item There are different possible paths leaving the point $a$ and leading to
different final points $b$, each having a probability $p_k(x|a)$ to be followed by the
system, where $x$ is the position of arbitrary $b$. This uncertainty is the basic
consideration for the definition of Kolmogorov-Sinai entropy\cite{Dorfman} in the
description of chaotic systems.

\end{enumerate}
In what follows, we first discuss the transition probability $p_k(b|a)$ between two
fixed phase points (or cells of a given partition of the phase space): the cell $a$ in
the initial phase volume $A$ and a cell $b$ in the final phase volume $B$ via a path
$k$ ($k=1,2,...w$). We suppose that the uncertainty concerning the choice of paths by
the systems between two points is measured with the following path information
\begin{eqnarray}                                            \label{c1x}
I_{ab}=-\sum_{k=1}^wp_k(b|a)\ln p_k(b|a).
\end{eqnarray}
We have the following normalization
\begin{eqnarray}                                            \label{c1xx}
\sum_{k=1}^{w}p_k(b|a)=1.
\end{eqnarray}

\section{Maximum path information}

An essential assumption of our informational method is that each path is characterized
by its action $A_{ab}(k)$ defined as if there was no random perturbation. This action
is given by
\begin{eqnarray}                                            \label{c5}
A_{ab}(k)=\int_{t_{ab}(k)}L_k(t)dt
\end{eqnarray}
where $L_{k}(t)=E-U$ is the Lagrangian of the system at time $t$ along the path $k$.
The average action is given by
\begin{eqnarray}                                            \label{c1xxx}
A_{ab}=\sum_{k=1}^wp_k(b|a)A_{ab}(k).
\end{eqnarray}
For irregular dynamic process, the least action principle does not apply since,
firstly, there is no exact calculation of real action due to dynamic randomness, and
secondly, the action we defined here has no stationary $\delta A_{ab}(k)=0$ for all
the possible paths. We use the information method which consists in saying that the
stable probability distribution of paths must correspond to a stationary uncertainty
or path information $I_{ab}$ under the constraint associated with the average action
$A_{ab}$. This means the following operation:
\begin{eqnarray}                                            \label{xc1x}
\delta [I_{ab}+\alpha\sum_{k=1}^{w}p_k(b|a)-\eta\sum_{k=1}^wp_k(b|a)A_{ab}(k)]=0
\end{eqnarray}
leading to
\begin{eqnarray}                                            \label{c6x}
p_k(b|a)=\frac{1}{Z}\exp[-\eta A_{ab}(k)],
\end{eqnarray}
where the partition function
\begin{eqnarray}                                            \label{cx6x}
Z=\sum_{k}\exp[-\eta A_{ab}(k)].
\end{eqnarray}
The physical meaning of the multiplier $\eta$ will be discussed below.

It is proved that\cite{Wang04x} the distribution Eq.(\ref{c6x}) is stable with respect
to the fluctuation of action. It is also proved that Eq.(\ref{c6x}) is a least
(stationary) action distribution, i.e., the most probable paths are just the paths of
least action. Indeed, we have $\delta p_k(b|a)=-\eta p_k(b|a)\delta A_{ab}(k)=0$ which
means $\delta A_{ab}(k)=0$ and yields Euler-Lagrange equation and Hamiltonian
equations given in Eqs.(\ref{1}) and (\ref{c7x}). Consequently, the most probable
paths are just the geodesics of the unperturbed hamiltonian dynamics. The other paths
do not satisfy Eqs.(\ref{1}) and (\ref{c7x}). In general, the paths have neither
$\delta A_{ab}(k)=0$ nor $\delta A_{ab}=0$. Nevertheless, the average action $A_{ab}$
defined over all these paths does have a stationary associated with the stationary of
the uncertainty $I_{ab}$ in Eq.(\ref{xc1x}), i.e.,
\begin{eqnarray}                                            \label{w6x}
-\eta\delta A_{ab}+\delta I_{ab}=0.
\end{eqnarray}
Here we considered $\sum_{k=1}^{w}\delta p_k(b|a)=0$. Eq.(\ref{w6x}) implies that,
although the least action principle of classical mechanics cannot apply when the
dynamics is perturbed by random and instable noise, the maximum path information
introduced above underlies in fact the same physics in which the action principle is
present as a average effect in association with the stationary dynamic uncertainty.
Eq.(\ref{w6x}) implies a probabilistic version of classical mechanics and will be used
below to derive the averaged Euler-lagrange equation for irregular dynamics.

\section{A calculation of least action distribution}
In what follows, the $unperturbed$ action is analyzed for different paths with the
Euler method. Let us look at a system of mass $m$ moving along a given path $k$ from a
point $a$ to a point $b$. The path is cut into $N$ infinitesimally small segments each
having a spatial length $\Delta x_i=x_i-x_{i-1}$ with $i=1 ...N$ ($x_0=x_a$ and
$x_N=x_b$). $t=t_i-t_{i-1}$ is the time interval spent by the system on every segment.
The Lagrangian on the segment $i$ is given by
\begin{eqnarray}                                            \label{x9c}
L(x_i,\dot{x}_i,t)=\frac{m(x_i-x_{i-1})^2}{2(t_i-t_{i-1})^2} -\left(\frac{\partial
U}{\partial x}\right)_i\frac{(x_i-x_{i-1})}{2}-U(x_{i-1})
\end{eqnarray}
where the first term on the right hand side is the kinetic energy of the particle, the
second and the third terms are the average potential energy on the segment $i$. The
action of segment $i$ is given by
\begin{eqnarray}                                            \label{xx9c}
A_i=\frac{m(\Delta x_i)^2}{2t} +F_i\frac{\Delta x_i}{2}t-U(x_{i-1})t,
\end{eqnarray}
where $F_i=-\left(\frac{\partial U}{\partial x}\right)_i$ is the force on the segment
$i$. According to Eq.(\ref{c6x}), the transition probability $p_{k,i}$ from $x_{i-1}$
to $x_i$ on the path $k$ is given by
\begin{eqnarray}                                            \label{xxxc9}
p_{k,i} &=& \frac{1}{Z_i} \exp\left(-\eta\left[\frac{m}{2t}\Delta x_i^2
+F_i\frac{t}{2}\Delta x_i\right]_{k}\right)
\end{eqnarray}
where $Z_i$ is calculated as follows
\begin{eqnarray}                                            \label{xxx9}
Z_i&=&\int_{-\infty}^\infty dx_i\exp\left(-\eta\left[\frac{m}{2t}\Delta x_i^2
+F_i\frac{t}{2}\Delta x_i\right]_{k}\right)\\ \nonumber &=& \exp\left[F_i^2\frac{\eta
t^3}{8m}\right]\sqrt{\frac{2\pi t}{m\eta}}.
\end{eqnarray}
The potential energy of the point $x_{i-1}$ disappears in the expression of
$p_{k,i}$ because it does not depend on $x_i$.

The total action is given by
\begin{eqnarray}                                            \label{cc9}
A_{ab}(k)=\sum_{i=1}^NA_i=\sum_{i=1}^N\left[\frac{m(\Delta x_i)^2}{2t}
+F_i\frac{t}{2}\Delta x_i-U(x_{i-1})t\right]_{k}.
\end{eqnarray}
According to Eq.(\ref{c6x}), the transition probability from $a$ to $b$ via the path
$k$ is the following:
\begin{eqnarray}                                            \label{c9}
p_k(b|a) &=& \frac{1}{Z} \exp\left(-\eta\sum_{i=1}^N\left[\frac{m(\Delta x_i)^2}{2t}
+F_i\frac{t}{2}\Delta x_i\right]_{k}\right)  \\ \nonumber &=&
p(b|a)^{-1}\prod_{i=1}^N p_{k,i}
\end{eqnarray}
where
\begin{eqnarray}                                            \label{c9x}
Z&=&\sum_{k=1}^{w}\exp\left(-\eta\sum_{i=1}^N\left[\frac{m(\Delta x_i)^2}{2t}
+F_i\frac{t}{2}\Delta x_i\right]_{k}\right) \\\nonumber &=&\int_{-\infty}^\infty
dx_1dx_2...dx_{N-1}\exp\left(-\eta\sum_{i=1}^N\left[\frac{m(x_i-x_{i-1})^2}{2t}
+F_i\frac{t}{2}(x_i-x_{i-1})\right]_{k}\right)\\\nonumber &=&
\left(\exp\left[F_i^2\frac{\eta t^3}{8m}\right]\sqrt{\frac{2\pi t}{m\eta}}\right)^N
p(b|a)=Z_i^Np(b|a)
\end{eqnarray}
and
\begin{eqnarray}                                            \label{ac9x}
p(b|a)&=& \exp\left[F_i^2\frac{\eta (t_b-t_a)^3}{8m}\right]\sqrt{\frac{m\eta}{2\pi
(t_b-t_a)}}
\\ \nonumber &\times&\exp\left(-\eta\left[\frac{m(x_b-x_a)^2}{2(t_b-t_a)}
+F_i\frac{t_b-t_a}{2}\Delta x_i\right]\right).
\end{eqnarray}
Remember that in the above calculation, the point $x_0=x_a$ and the final point
$x_N=x_b$ are fixed.

Now in order to see the behavior of transition probability with respect to final
point, we have to relax $x_b=x$ and let it vary arbitrarily as other intermediate
points. This implies we take into account the second uncertainty due to chaos
mentioned in the introduction. The corresponding transition probability $p_k(x|a)$
from $a$ to arbitrary $x$ via the path $k$ has been derived with the maximum path
information combined with action\cite{Wang04xx}. Here we only introduce it in the
following way:
\begin{eqnarray}                                            \label{acc9x}
p_k(x|a) &=& p(b|a)p_k(b|a) \\
\nonumber &=& \prod_{i=1}^N p_{k,i/i-1},
\end{eqnarray}
normalized by
\begin{eqnarray}                                            \label{acc1xx}
\sum_b\sum_{k=1}^{w}p_k(x_b|a)=\int dx_1dx_2...dx_{N-1}dxp_k(x|a)=1.
\end{eqnarray}
The uncertainty or path information associated with this transition probability
between $a$ and arbitrary $b$ is given by

\begin{eqnarray}                                            \label{c1w}
I_{a}&=&-\sum_b\sum_{k=1}^wp_k(x|a)\ln p_k(x|a)\\ \nonumber &=&-\sum_bp(b|a)\ln
p(b|a)-\sum_bp(b|a)\sum_{k=1}^wp_k(b|a)\ln p_k(b|a)\\ \nonumber &=& h_{ab}+\langle
I_{ab}\rangle
\end{eqnarray}
where $h_{ab}=-\sum_bp(b|a)\ln p(b|a)$ is the uncertainty of the transition from $a$
to an arbitrary $b$ via whichever path, and $\langle I_{ab}\rangle$ is the average of
$I_{ab}$ over all the possible point $b$ in the final phase volume $B$.

The average action defined with $p_k(x|a)$ over all the possible paths and final
points is given by
\begin{eqnarray}                                            \label{w1w}
\langle A_{ab}\rangle&=&-\sum_b\sum_{k=1}^wp_k(x|a)A_{ab}(k)\\ \nonumber
&=&\sum_bp(b|a)\sum_{k=1}^wp_k(b|a)A_{ab}(k)\\ \nonumber &=& \sum_bp(b|a)A_{ab}.
\end{eqnarray}
Remember that $p(b|a)$ does not depend on the paths. So if $A_{ab}$ has a stationary,
$\langle A_{ab}\rangle$ must have a stationary as well.

\section{Fokker-Planck equation}
A derivation of the diffusion laws from Eq.(\ref{xxxc9}) and Eq.(\ref{acc9x}) is given
in \cite{Wang04b}. In what follows, we only illustrate this straightforward method
with Fokker-Planck equation which describes the time evolution of transition
probability. This equation can be derived if we suppose that the diffusion particles
follow Brownian motion and Markovian process\cite{Kubo}, or Kolmogorov
conditions\cite{Zaslavsky}.

From Eq.(\ref{xxxc9}), a simple calculation of the derivatives $\frac{\partial
p_{i}}{\partial t}$, $\frac{\partial (F_ip_{i})}{\partial x_i}$ and $\frac{\partial^2
p_{i}}{\partial x_i^2}$ yields
\begin{eqnarray}                                            \label{xc10}
\frac{\partial p_i}{\partial t} = -\frac{\tau}{m}\frac{\partial (F_ip_i)}{\partial
x_i}+\frac{1}{2m\eta}\frac{\partial^2 p_{i}}{\partial x_i^2}.
\end{eqnarray}
This is the Fokker-Planck equation, where $\tau$ is the mean free time supposed to be
the time interval $t$ of the particle on each segment of its path. In view of the
Eq.(\ref{acc9x}), it is easy to show that this equation is also satisfied by
$p_k(x|a)$ if $x_i$ is replaced by $x$, the final position. Other diffusion laws can
be easily obtained from the above equation. We think that this derivation is a proof
of the usefulness of the present informational approach to irregular dynamics.

As discussed in \cite{Wang04b}, Eq.(\ref{xc10}) yields in general:
\begin{eqnarray}                                            \label{w10}
\frac{\partial n}{\partial t} = -\frac{\tau}{m}\frac{\partial (nF)}{\partial
x}+\frac{1}{2m\eta}\frac{\partial^2 n}{\partial x^2}.
\end{eqnarray}
where $n$ is the particle density at point $x$ and time $t$. In order to clarify the
physical meaning of $\eta$, we consider here a gas in equilibrium (hence
$\frac{\partial n}{\partial t}=0$) in a constant force $F$ with the distribution
$n(x)=n(x_0)exp[F(x-x_0)/k_BT]$ where $k_B$ is the Boltzmann constant. From
Eq.(\ref{w10}), we straightforwardly get
\begin{eqnarray}                                            \label{w9v}
\eta = \frac{1}{2\tau k_BT}
\end{eqnarray}

Now let us focus on the modification imposed upon the formalism of classical
mechanics by the above probabilistic description of stochastic mechanics.

\section{Euler-Lagrange equations}
Let us look at, not the most probable paths (geodesics) of the dynamics from $A$ to
$B$ which satisfy Eqs.(\ref{1}) and (\ref{c7x}), but all the other paths whose
$unperturbed$ action is not a stationary. When a system travels along these paths,
$\delta A_{ab}(k) \neq 0$. It is either positive or negative. On the other hand, we
have\cite{Arnold}
\begin{eqnarray}                                            \label{7aa}
\delta A_{ab}(k) &=& \int_a^b \left[\frac{\partial}{\partial t}\frac{\partial
L_{k}(t)}{\partial \dot{x}}-\frac{\partial L_{k}(t)}{\partial x}\right]\varepsilon dt
\end{eqnarray}
where $\varepsilon$ is an arbitrary variation of $x$ ($\varepsilon$ is zero at $a$ and
$b$). If $\delta A_{ab}(k) > 0$ (or $< 0$), we get
\begin{eqnarray}                                            \label{7ab}
\frac{\partial}{\partial t}\frac{\partial L_{k}(t)}{\partial \dot{x}}-\frac{\partial
L_{k}(t)}{\partial x}> 0\;\;(or <0).
\end{eqnarray}
which can be proved as follows. Suppose $\int_a^b f(t)\varepsilon dt>0$ and
$f(t)<c<0$ during a small period of time $\Delta t$ somewhere between $a$ and $b$.
Since $\varepsilon$ is arbitrary, let it be zero outside $\Delta t$ and a positive
constant within $\Delta t$. We clearly have $\int_a^b f(t)\varepsilon
dt<c\varepsilon<0$ which contradicts our starting assumption. This proves
Eq.(\ref{7ab}).

Legendre transformation $H=P\dot{x}-L_{k}$ implies\cite{Arnold}
\begin{eqnarray}                                            \label{8a}
P=\frac{\partial L_{k_{ab}}}{\partial \dot{x}}.
\end{eqnarray}
Put this relationship back into Eq.(\ref{7ab}), we get
\begin{eqnarray}                                            \label{8b}
\dot{P}> \;(or <)\;\frac{\partial L_{k}}{\partial x}
\end{eqnarray}
for $dA_{ab}(k) > 0$ (or $< 0$). In view of the fact that the deviation of $\dot{P}$
from the regular force $\frac{\partial L_{k}}{\partial x}$ is due to the dynamic
irregularity, by introducing a ``force'' $R$ of the random perturbation,
Eq.(\ref{8b}) can be recast into
\begin{eqnarray}                                            \label{8bb}
\dot{P}=\frac{\partial L_{k}}{\partial x}+R.
\end{eqnarray}
According to Eq.(\ref{8b}), $R$ is positive if $dA_{ab}(k)>0$ and negative if $d
A_{ab}(k)<0$.

\section{Hamiltonian equations}
The total differential of $H$ is
\begin{eqnarray}                                            \label{8c}
dH=\frac{\partial H}{\partial P}dP+\frac{\partial H}{\partial x}dx +\frac{\partial
H}{\partial t}dt.
\end{eqnarray}
From Legendre transformation, we can also write\cite{Arnold}
\begin{eqnarray}                                            \label{8d}
dH=\dot{x}dP-\frac{\partial L_{k}}{\partial x}dx -\frac{\partial L_{k}}{\partial
t}dt.
\end{eqnarray}
Comparing Eq.(\ref{8c}) to Eq.(\ref{8d}), we get
\begin{eqnarray}                                            \label{1a}
\dot{x}=\frac{\partial H}{\partial P} \;\;and \;\; \dot{P}> \;(or <)\;
-\frac{\partial H}{\partial x},
\end{eqnarray}
which are the Hamiltonian equations for irregular dynamic process following the
paths whose action is not at stationary. The second equation above violates the
Newton's second law. Remember that $>$ (or $<$) is for the case where
$dA_{ab}(k)=\frac{\partial A_{ab}(k)}{\partial x}dx -\frac{\partial
A_{ab}(k)}{\partial \dot{x}}d\dot{x}>\;(or <)\;0$.

Considering the random force $R$, Eq.(\ref{1a}) can be written as
\begin{eqnarray}                                            \label{wcx6x}
\dot{P}_k=-\frac{\partial H}{\partial x}+R.
\end{eqnarray}
This is a Langevin equation.
\section{Energy conservation}
From Eq.(\ref{8c}), we see that if the Hamiltonian of the system is not an explicit
function of time ($\frac{\partial H}{\partial t}=0$), we get
\begin{eqnarray}                                            \label{w8c}
\frac{dH}{dt}=\frac{\partial H}{\partial P}\dot{P}+\frac{\partial H}{\partial
x}\dot{x}.
\end{eqnarray}
For the geodesics, using Eq.(\ref{1}), we get $\frac{dH}{dt}=0$. So the energy is
conserved during the time evolution of the system. But if the system travels along
other paths than the unperturbed geodesics, Eq.(\ref{1a}) occurs, we should write
$\frac{dH}{dt}> \;(or <)\;0$ for the paths having $dA_{ab}(k) > 0$ (or $< 0$). The
energy is no more conserved.

The fluctuation of energy is defined by $\sigma_H^2=\langle(H_k-\langle
H_k\rangle)^2\rangle=\langle H_k^2\rangle-\langle H_k\rangle^2$ which we will
calculate for a segment $i$ of the paths in the time interval $\tau=t_i-t_{i-1}$
(see section 5). The action on this segment of the path $k$ can be written as
\begin{eqnarray}                                            \label{w9}
A_i=(P_i\dot{x}_i-H_i)_k\tau
\end{eqnarray}
So the transition probability is given by
\begin{eqnarray}                                            \label{wc9}
p_{k,i} &=& \frac{1}{Z_i} \exp\left[-\eta\tau(P_i\dot{x}_i-H_i)_k\right]
\end{eqnarray}
where $Z_i=\sum_k\exp\left[-\eta\tau(P_i\dot{x}_i-H_i)_k\right]$. After some
mathematics, we get
\begin{eqnarray}                                            \label{ww9}
\sigma_H^2 &=& \frac{1}{Z_i} \sum_k(H_i-\langle H_i\rangle)^2\exp\left[-\eta
\tau(P_i\dot{x}_i-H_i)_k\right] \\\nonumber &=& \frac{1}{\tau}\frac{\partial \langle
H_i\rangle}{\partial \eta}+\langle P\dot{x}H_i\rangle-\langle P\dot{x}\rangle\langle
H_i\rangle.
\end{eqnarray}

Let's see an example with {\it free particle}. We have in this case $\langle
P\dot{x}H_i\rangle=2\langle H_i^2\rangle$ and $\langle P\dot{x}\rangle\langle
H_i\rangle=2\langle H_i\rangle^2$, leading to
\begin{eqnarray}                                            \label{xw9}
\sigma_H^2 &=& -\frac{1}{\tau}\frac{\partial \langle H_i\rangle}{\partial \eta}.
\end{eqnarray}
For this kind of systems, $\langle H_i\rangle$ can be calculated by using
Eq.(\ref{xxxc9}) with $F_i=0$. We have $\langle\Delta
x_i^2\rangle=\frac{\tau}{m\eta}$ an $\langle H_i\rangle=\frac{m\langle\Delta
x_i^2\rangle}{2\tau^2}=\frac{1}{2\tau\eta}=k_BT$. Finally, we get
\begin{eqnarray}                                            \label{xw9w}
\sigma_H^2 = \frac{1}{2\tau^2\eta^2}=2k_B^2T^2.
\end{eqnarray}
Don't forget that we are addressing one dimensional system which can have only one
particle under thermal noise. For $N$-particle systems, the result is different.
From Eqs.(\ref{w9v}) and (\ref{xw9}), one can write

\begin{eqnarray}                                            \label{vw9}
\sigma_H^2 &=& -2C_ik_BT^2
\end{eqnarray}
where $C_i=\frac{\partial \langle H_i\rangle}{\partial T}$ is the heat capacity of
the system on the segment $i$ of the paths.

\section{Liouville theorem}

Now we look at the time change of state density $\rho(\Gamma)$ in phase space. The
time evolution neither creates nor destroys state points, so the number of the phase
points is conserved. The conservation law in a 2-dimensional phase space is
\begin{eqnarray}                                            \label{q7a}
\frac{\partial \rho}{\partial t}+ \frac{\partial (\dot{x}\rho)}{\partial x}+
\frac{\partial (\dot{P}\rho)}{\partial P}=0
\end{eqnarray}
which means
\begin{eqnarray}                                            \label{q7b}
\frac{d\rho}{dt}=\frac{\partial \rho}{\partial t}+ \frac{\partial \rho}{\partial
x}\dot{x}+ \frac{\partial \rho}{\partial P}\dot{P}=-(\frac{\partial\dot{x}}{\partial
x}+ \frac{\partial\dot{P}}{\partial P})\rho.
\end{eqnarray}
For the least action paths which satisfy Eqs.(\ref{1}), the right hand side of the
above equation is zero, leading to the {\it Liouville theorem}
\begin{eqnarray}                                            \label{x7b}
\frac{d\rho(\Gamma,t)}{dt}=0,
\end{eqnarray}
i.e., the state density in phase space is a constant of motion.

For a system travelling along a perturbed paths having $dA_{ab}(k)\neq 0$,
Eqs.(\ref{1}) vanishes and so does Eq.(\ref{x7b}). Considering Eqs.(\ref{wcx6x}) and
(\ref{q7b}), we find
\begin{eqnarray}                                            \label{8}
\frac{d\rho(\Gamma,t)}{dt} = -\frac{\partial R}{\partial P}\rho.
\end{eqnarray}
Nevertheless, the determination of the sign of $\frac{\partial R}{\partial P}$ is
not evident in general. In any case, this result means that, in general, if a system
does not travel along the paths of stationary $unperturbed$ action due to random or
molecular noise, then its state density may change with time.

The first consequence of Eq.(\ref{8}) is that, in general, the phase volume $\Omega$
is no more conserved for irregular dynamic systems since
\begin{eqnarray}                                            \label{9}
\frac{d\Omega}{dt}=\frac{d}{dt}\int\rho(\Gamma,t)d\Gamma
=\int\frac{d\rho(\Gamma,t)}{dt}d\Gamma=-\overline{\frac{\partial R}{\partial P}}\neq
0,
\end{eqnarray}
where the average $\overline{\frac{\partial R}{\partial P}}$ is over all the phase
volume occupied by the system at time $t$.

\section{Averaged formalism of stochastic mechanics}
\subsection{Euler-Lagrange equations}
It is mentioned above that the dynamics described by the distribution Eq.(\ref{c6x})
has a {\it stationary average action} in association with the stationary path
information. In what follows, we use this fact to derive the averaged version of the
equations of motion of classical mechanics.

$\delta A_{ab}$ can be calculated as follows:
\begin{eqnarray}                                            \label{7}
\delta A_{ab} &=& \sum p_k(b|a)\delta A_{ab}(k)+\sum \delta p_k(b|a)A_{ab}(k).
\end{eqnarray}

The first term on the right hand side of this equation is
\begin{eqnarray}                                            \label{7a}
\sum p_k(b|a)\delta A_{ab}(k) &=& \sum p_k(b|a)\int_a^b
\left[\frac{\partial}{\partial t}\frac{\partial L_{k}(t)}{\partial
\dot{x}}-\frac{\partial L_{k}(t)}{\partial x}\right]\varepsilon dt \\
\nonumber &=& \int_a^b \left[\left\langle\frac{\partial}{\partial t}\frac{\partial
L_{k}(t)}{\partial \dot{x}}\right\rangle-\left\langle\frac{\partial
L_{k}(t)}{\partial x}\right\rangle\right]\varepsilon dt
\end{eqnarray}
where $\left\langle\cdot\right\rangle$ is the average over all the possible paths.

The second term on the right hand side of Eq.(\ref{7}) is
\begin{eqnarray}                                            \label{7b}
\sum \delta p_k(b|a)A_{ab}(k)&=&-\frac{1}{\eta}\sum \delta p_k(b|a)\ln[Zp_k(b|a)]
\\ &=&-\frac{1}{\eta}\delta\sum p_k(b|a)\ln p_k(b|a)=\frac{\delta I_{ab}}{\eta}.
\end{eqnarray}
where we have used $\sum \delta p_k(b|a)=0$. Considering Eq.(\ref{w6x}), we get
\begin{eqnarray}                                            \label{7c}
\int_a^b \left[\left\langle\frac{\partial}{\partial t}\frac{\partial
L_{k}(t)}{\partial \dot{x}}\right\rangle-\left\langle\frac{\partial L_{k}(t)}{\partial
x}\right\rangle\right]\varepsilon dt=\delta A_{ab}-\frac{\delta I_{ab}}{\eta}=0.
\end{eqnarray}
Now considering that $\varepsilon$ is arbitrarily chosen, we obtain
\begin{eqnarray}                                            \label{7d}
\left\langle\frac{\partial}{\partial t}\frac{\partial L_{k}(t)}{\partial
\dot{x}}\right\rangle-\left\langle\frac{\partial L_{k}(t)}{\partial
x}\right\rangle=0.
\end{eqnarray}
This is the {\it averaged Euler-Lagrange equation}.

From Eq.(\ref{8a}), we can write
\begin{eqnarray}                                            \label{8ab}
\left\langle\dot{P}\right\rangle=\left\langle\frac{\partial L_{k}(t)}{\partial
x}\right\rangle
\end{eqnarray}
where $\left\langle\frac{\partial L_{k}(t)}{\partial x}\right\rangle$ is an averaged
force.

\subsection{Averaged Hamiltonian equations}
By means of Eq.(\ref{1a}) and $\frac{\partial H(t)}{\partial x}=-\frac{\partial
L_{k_{ab}}(t)}{\partial x}$, the following averaged equations exist

\begin{eqnarray}                                            \label{1ab}
\left\langle\dot{x}\right\rangle=\left\langle\frac{\partial H}{\partial
P}\right\rangle \;\;and \;\;
\left\langle\dot{P}\right\rangle=-\left\langle\frac{\partial H}{\partial
x}\right\rangle.
\end{eqnarray}
This implies that the mean of the ``random force'' $R$ over all possible paths must
vanish, i.e., $\left\langle R\right\rangle=0$.

\subsection{Energy conservation law}
From the energy conservation problem discussed above, we see that the only
Hamiltonian which is physically significant for randomly perturbed hamiltonian
systems is the average Hamiltonian $\langle H\rangle$. Its derivative with respect
to time (change rate of average energy) is
\begin{eqnarray}                                            \label{w8w}
\frac{d\langle H\rangle}{dt}=\langle\frac{dH}{dt}\rangle+\sum_k\frac{dp_{k,i}}{dt}H.
\end{eqnarray}
The first term of this equation is just
\begin{eqnarray}                                            \label{w8wx}
\langle\frac{dH}{dt}\rangle=\langle\frac{\partial H}{\partial P}\dot{P}\rangle
+\langle\frac{\partial H}{\partial x}\dot{x}\rangle=\langle
\dot{x}R\rangle=\langle\frac{dW_R}{dt}\rangle
\end{eqnarray}
where $W_R$ is a random work done by the ``random force'' $R$.

With the help of the least action distribution $p_{k,i}$ and the definition of path
information, the second term of Eq.(\ref{w8w}) can be given by
\begin{eqnarray}                                            \label{w9wx}
\sum_k\frac{dp_{k,i}}{dt}H=\sum_k\frac{dp_{k,i}}{dt}P\dot{x}-\frac{1}{\tau\eta}\frac{d
I_{i}}{dt}
\end{eqnarray}
where $I_{i}=-\sum_kp_{k,i}\ln p_{k,i}$ is the path information on the segment $i$
and in the time interval $\tau$. Simple calculation leads to
\begin{eqnarray}                                            \label{d9wx}
\frac{d I_{i}}{dt}=\eta\left[\frac{d \langle A_{i}\rangle}{dt}-\left\langle\frac{d
A_{i}}{dt}\right\rangle\right]=\eta(\Delta_tA)
\end{eqnarray}
On the other hand,
\begin{eqnarray}                                            \label{s10wx}
\sum_k\frac{dp_{k,i}}{dt}P\dot{x}&=&\frac{d\langle
P\dot{x}\rangle}{dt}-\sum_kp_{k,i}\frac{d(P\dot{x})}{dt}\\
\nonumber &=&\frac{d\langle P\dot{x}\rangle}{dt}-2\sum_kp_{k,i}[-\frac{\partial
H}{\partial x}+R]\dot{x}\\ \nonumber &=& \frac{d\langle
P\dot{x}\rangle}{dt}+2\langle\frac{\partial H}{\partial x}\dot{x}\rangle-2\langle
\dot{x}R\rangle.
\end{eqnarray}
Hence we get
\begin{eqnarray}                                            \label{10wx}
\sum_k\frac{dp_{k,i}}{dt}H&=&\frac{d\langle
P\dot{x}\rangle}{dt}+2\langle\frac{\partial H}{\partial x}\dot{x}\rangle-2\langle
\dot{x}R\rangle-\frac{1}{\tau}(\Delta_tA).
\end{eqnarray}
This means
\begin{eqnarray}                                            \label{11wx}
\frac{d\langle H\rangle}{dt}&=& \frac{d\langle
P\dot{x}\rangle}{dt}+2\langle\frac{\partial H}{\partial x}\dot{x}\rangle-\langle
\dot{x}R\rangle-\frac{1}{\tau}(\Delta_tA)
\\ \nonumber &=&2\frac{d\langle T\rangle}{dt}-2\langle\frac{dW_H}{dt}\rangle-\langle
\dot{x}R\rangle-\frac{1}{\tau}(\Delta_tA)
\end{eqnarray}
where $T$ is the kinetic energy of the system, $W_H$ is the work done by the
conservative force. We suppose that our hamiltonian system neither gains nor loses
energy through the irregular random process. So the average energy must be a
constant of motion, i.e., $\frac{d\langle H\rangle}{dt}=0$. This means the random
work must satisfy the following relation:
\begin{eqnarray}                                            \label{12}
\langle \dot{x}R\rangle=2\left(\frac{d\langle
T\rangle}{dt}-\left\langle\frac{dW_H}{dt}\right\rangle\right)-\frac{1}{\tau}(\Delta_tA).
\end{eqnarray}
This condition seems reasonable because a random force, though statistically
vanishing, can made non vanishing average work which may change the average energy
of a system. So if we require that the system statistically remain hamiltonian, this
work or its power $\langle \dot{x}R\rangle$ should satisfy some condition. If ever
we have $\langle \dot{x}R\rangle=0$, we can write $\langle\frac{dH}{dt}\rangle=0$,
meaning that the time variation of energy is statistically null. But the average
energy may always change in time due to the deviation from the least action paths.
We have the following condition for statistical energy conservation:
\begin{eqnarray}                                            \label{12x}
\left(\frac{d\langle
T\rangle}{dt}-\left\langle\frac{dW_H}{dt}\right\rangle\right)=\frac{1}{2\tau}(\Delta_tA).
\end{eqnarray}

\section{Passage from stochastic mechanics to regular one}
Of course, when the randomness diminishes, this stochastic mechanics formalism
should recover the regular classical one. In fact, the dispersion of action or the
width of the least action distribution can be measured by the variance
$\sigma^2=\overline{A^2}-\overline{A}^2=\overline{A^2}-A_{ab}^2$. From
Eqs.(\ref{c1x}), (\ref{c1xxx}), (\ref{c6x}) and (\ref{cx6x}), we get
\begin{eqnarray}                                            \label{w12}
\sigma^2=-\frac{\partial A_{ab}}{\partial \eta},
\end{eqnarray}

\begin{eqnarray}                                            \label{w7x}
A_{ab}=-\frac{\partial}{\partial\eta}\ln Z,
\end{eqnarray}
and
\begin{eqnarray}                                            \label{wc7}
I_{ab}=\ln Z+\eta A_{ab}=\ln Z-\eta \frac{\partial}{\partial\eta}\ln Z
\end{eqnarray}
When the system becomes less and less irregular, $\sigma^2$ diminishes and the paths
become closer and closer to the least action ones having stationary action
$A_{ab}^{stat}$. When $\sigma^2\rightarrow 0$, the significant contribution to the
partition function $Z$ comes from the geodesics having $A_{ab}^{stat}$, i.e.,
$Z\rightarrow exp[-\eta A_{ab}^{stat}]$. From Eq.(\ref{w7x}), $A_{ab}\rightarrow
A_{ab}^{stat}$. Then considering Eq.(\ref{wc7}), it is clear that the path
information $I_{ab}\rightarrow 0$. In this case, the stationary average action
Eq.(\ref{w6x}) becomes $\delta A_{ab}^{stat}=0$, the usual action principle, and
Eqs.(\ref{7ab}) and (\ref{1a}) will recover Eqs.(\ref{1}) and (\ref{c7x}). At the
same time, diffusion phenomena completely vanish and the diffusion laws are replaced
by the laws of regular mechanics. The energy conservation law is also recovered
because  Eq.(\ref{12x}) becomes $\left(\frac{dT}{dt}=\frac{dW_H}{dt}\right)$ due to
$\Delta_tA=\frac{d \langle A_{i}\rangle}{dt}-\left\langle\frac{d
A_{i}}{dt}\right\rangle=0$.

\section{Concluding remarks}
The objective of this paper is to present certain mathematical aspects of an
information method formulated in the attempt to describe diffusion phenomena within
a formalism of stochastic mechanics. We address hamiltonian systems undergoing
irregular dynamic process due to random perturbation of thermal (molecular) noise
and chaotic instability. The dynamic randomness or irregularity is taken into
account via the uncertainty associated with different paths between two points in
phase space and with the different final points from a given initial point. These
uncertainties are measured by path informations maximized in connection with average
action. It results that the probability of the irregular process depends
exponentially on action (least action distribution). The usefulness of this
information method is demonstrated by the derivation of diffusion laws such as the
Fokker-Planck equation, Fick's laws, Ohm's law\cite{Wang04b} and the Fourier's
law\cite{Wang04bx}.

This information approach with the least action distribution of paths implies a
deviation from the classical regular dynamics and a formalism of stochastic
mechanics. The instantaneous behaviors of irregular hamiltonian system may violate
classical mechanics and its fundamental equations if only the unperturbed
Hamiltonian is introduced. This is understandable because, due to the random noise
and instability, exact mathematical treatment of the perturbed Hamiltonian is
inconceivable even when the system remain statistically hamiltonian. In this case,
only the average Hamiltonian is physically significant.

Another point is that this approach underlines a tight connection between the
diffusion laws and the maximum dynamic uncertainty associated with the average
action. Remember that the action here is that of the unperturbed regular system, not
the real action (does it exist?) of the randomly perturbed system. This association
of dynamic uncertainty with average action is in fact an application of the action
principle to the whole dynamic situation, i.e., all the paths, not only the most
probable paths or geodesics. As a consequence, it is the average action, instead of
action, which has a stationary under the constraint of uncertainty.

This averaged action principle underlies a formalism of stochastic mechanics, which
can be derived directly from the violated fundamental equations by probabilistic
consideration. The mathematical form of classical mechanics can be recovered. The
fundamental equations are statistically satisfied. It is shown that if the work of
``random force'' satisfy some condition like Eq.(\ref{7}), the system remains
Hamiltonian with a constant average energy. We would like to emphasize here that the
Liouville theorem is also violated by the ensemble of paths or by the instantaneous
behaviors of random hamiltonian system. But it seems not obvious to recover this
theorem within the stochastic version of classical mechanics. This result may have
impacts on some properties of mechanical systems, e.g., on the Poincar\'e recurrence
theorem and the mechanical interpretation of the second law of thermodynamics.
Further work is in progress to tackle this topic.

\end{document}